\documentclass[11pt,twoside]{article}
\usepackage{amsmath,amssymb}
\numberwithin{equation}{section}

\begin{document}
\date{}
\title{\textbf{6-Body Central Configurations Formed  by Tow Isosceles Triangles}\footnote{This work is partially supported by NSF of China and Youth found of Mianyang Normal University.}}

\maketitle
\indent \indent \indent  \indent \indent \indent  \author{Furong Zhao$^{1,2}$\, and Shiqing Zhang$^1$ }

\begin{center}
$^1$Department of Mathematics, Sichuan University, Chengdu, 610064,P.R.China\\
$^{2}$Department of Mathematics and Computer Science, Mianyang Normal
University,
Mianyang, Sichuan,621000,P.R.China\\
\end{center}

\textbf{Abstract}:
In this paper,we show the existence of a class of 6-body central configurations with two isosceles triangles;which are congruent to each other and keep some distance.We also study the necessary conditions about masses for the bodies which can form  a central configuration.
\baselineskip=20pt
\section{Introduction and Main Results }

\indent \indent The Newtonian N-body problem  concerns the motion of N particles with masses $m_j \in R^{+}$ and positions  $q_j \in R^{3}$$(j=1,2,... ,N)$ ,the motion  is governed by Newton's second law and the Universal law:\\
\begin{equation}
m_{j}\ddot{q}_{j}=\frac{\partial U(q)}{\partial {q}_{j}},
\end{equation} where $q=(q_{1},q_{2},\cdots,q_{N})$ and $U(q)$ is Newtonian potential:\\
\begin{equation}
U(q)=\sum_{1\leqslant j<k\leqslant N }\frac{m_{j}m_{k}}{|q_{j}-q_{k}|},
\end{equation}
Consider the space\\
\begin{equation}
X=\{q=(q_{1},q_{2},\cdots,q_{N})\in R^{3N}:\sum_{j=1}^{N}m_{j}q_{j}=0 \},
\end{equation}
i.e,suppose that the center of mass is fixed at the origin of the space.
Because the potential is singular when two particles have same position,
it is natural to assume that the configuration avoids the collision set
$\triangle =\{q=(q_1,\cdots,q_N):q_j=q_k $ for some $k\neq j\}$.The set $X\backslash \triangle $
is called the configuration
space.\\
\indent \textbf{Definition 1.1}([17,22]):A configuration $q=(q_{1},q_{2},\cdots,q_{N})\in X\backslash \triangle$
is called a central configuration if there exists a constant $\lambda$ such that\\
\begin{equation}
\sum_{j=1,j\neq k}^{N}\frac{m_{j}m_{k}}{|q_{j}-q_{k}|^{3}}(q_{j}-q_{k})=-\lambda m_{k}q_{k},1\leqslant k\leqslant N.
\end{equation}
The value of constant $\lambda$ in (1.4) is uniquely determined by \\
\begin{equation}
\lambda=\frac{U}{I},
\end{equation}
where
\begin{equation}
I=\sum_{k=1}^{N}m_{k}|q_{k}|^{2}.
\end{equation}
\indent  Since the general solution of the N-body problem can't be
given, great importance has been attached to search for particular solutions from the
 very beginning. A homographic solution is that a configuration is preserved for all time.
 Central configurations and homographic solutions are linked by the Laplace theorem ([17,22]).
 Collapse orbits and parabolic orbits have relations with
the central configurations([15,16]).So finding central
configurations becomes very important. The main general open
problem for the cental configurations is due to Winter[22]and
Smale[20]:Is the number of planar central configurations finite
for any choice of positive masses $m_1,...,m_N $?Hampton and Moeckel([6])
have proved this conjecture for four any  given positive masses. \\
    In 1941, Wintner([22]) have studied regular polygon central configurations.  \\
Moeckel ([11]),Zhang and Zhou([23]) have studied highly symmetrical central configuration of Newtonian N-body problems.Llibre and Mello ([8]) have studied
a class of 6-body central configurations.\\
Based the above works,we find a classes of central configurations in the 6-body problems, for which  three bodies are at the vertices
of an isosceles triangles , the others are located at the vertices of another isosceles triangles and the two triangles are congruent to each other;\\
    Related assumptions will be interpreted more precisely in the following.\\
    Assume $m_1,m_2$ and $m_5$  are located the vertices
of a isosceles triangles $\Delta_1$;$m_3,m_4$ and $m_6$ are located at the vertices
of another isosceles triangles $\Delta_2$.
$\Delta_1$ and $\Delta_2$ are coplanar and  are congruent to each other;$q_1-q_2$ is parallel to $q_4-q_3$;$|q_1-q_4|<|q_5-q_6|$;$q_5$ and $q_6$ are located at the common perpendicular bisector for $q_1q_2$ and $q_3q_4$.Without loss of generality we can take a coordinate system such that
$q_1=(-1,y)$,$q_2=(-1,-y)$,$q_3=(1,-y)$,
$q_4=(1,y)$,$q_5=(-1-x,0)$,$q_6=(1+x,0)$.(See Fig).\\
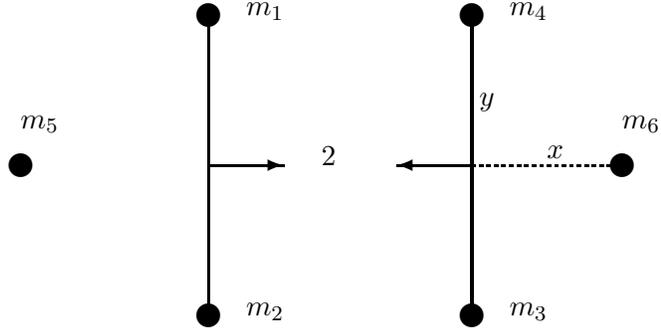
\begin{center}
\setlength{\unitlength}{1mm}
\begin{picture}(80,40)
\thicklines

\put(30,40){$m_1$}
\put(30,0){$m_2$}
\put(0,25){$m_5$}
\put(25,40){\circle*{3}}
\put(25,0){\circle*{3}}
\put(0,20){\circle*{3}}

\put(80,25){$m_6$}
\put(80,20){\circle*{3}}
\put(65,0){$m_3$}
\put(60,0){\circle*{3}}

\put(65,40){$m_4$}
\put(60,40){\circle*{3}}

\put(25,0){\line(0,1){40}}
\put(60,0){\line(0,1){40}}
\put(25,20){\vector(1,0){10}}
\put(60,20){\vector(-1,0){10}}
\multiput(60,20)(1,0){20}{\line(1,0){0.5}}
\put(70,21){$x$}
\put(61,28){$y$}
\put(40,20){2}
\end{picture}
\end{center}
We have:\\
\textbf{Theorem1.1}:If  $m_1,m_2,m_3,m_4,m_5$ and $m_6$ form a central configuration,then $m_1=m_2=m_3=m_4$ and $m_5=m_6$.\\
\textbf{Theorem1.2}:Assume $m_1=m_2=m_3=m_4=1$,$m_5=m_6=m$,
then  there exists exists a non-empty open set $U\subset(1,+\infty),$
$\varphi(y)\in C(U)$ such that $\varphi(\sqrt{3})=1$ and $m=m(x,y)$=$m(\varphi(y),y)$, so that
$(q_1,q_2,q_3,q_4,q_5,q_6)$ form a central configuration.\\
\textbf{Remark}:When $x=1$ and $y=\sqrt{3}$, $q_i$ is the vertex of a regular 6-gons $(i=1,\cdots,6)$.

\section{The Proofs of Theorems }
\subsection{The Proof of Theorem 1.1}
Note that
\begin{equation}
\begin{split}
\sum_{j=1,j\neq k}^{N}\frac{m_{j}m_{k}}{|q_{j}-q_{k}|^{3}}(q_{j}-q_{k})=-\lambda m_{k}q_{k}=-\lambda m_{k}(q_{k}-0)\\=
-\lambda m_{k}(q_{k}-\frac{\sum_{j=1}^{N}m_jq_j}{M})
=-m_{k}\frac{\lambda}{M} \sum_{j=1}^{N}m_j(q_{k}-q_j)
\end{split}
\end{equation}
where $M=\sum_{i=1}^Nm_i$.\\
So (1.4) is also  equivalent to
\begin{equation}
\sum_{j=1,j\neq k}^{N}m_{j}(\frac{1}{|q_{j}-q_{k}|^{3}}-\frac{\lambda}{M})(q_{j}-q_{k})=0
\end{equation}
By (2.2) we have
\begin{equation}
\begin{split}
0m_1+0m_2+2(\frac{1}{|4+4y^2|^{3/2}}-\frac{\lambda}{M})m_3+2(\frac{1}{2^3}-\frac{\lambda}{M})m_4\\
-x(\frac{1}{|x^2+y^2|^{3/2}}-\frac{\lambda}{M})m_5+(2+x)(\frac{1}{|x^2+y^2+4x+4|^{3/2}}-\frac{\lambda}{M})m_6=0,
\end{split}
\end{equation}
\begin{equation}
\begin{split}
0m_1-2y(\frac{1}{|2y|^3}-\frac{\lambda}{M})m_2-2y(\frac{1}{|4+4y^2|^{3/2}}-\frac{\lambda}{M})m_3+0m_4\\
-y(\frac{1}{|x^2+y^2|^{3/2}}-\frac{\lambda}{M})m_5-y(\frac{1}{|x^2+y^2+4x+4|^{3/2}}-\frac{\lambda}{M})m_6=0,
\end{split}
\end{equation}
\begin{equation}
\begin{split}
0m_1+0m_2+2(\frac{1}{2^3}-\frac{\lambda}{M})m_3+2(\frac{1}{|4+4y^2|^{3/2}}-\frac{\lambda}{M})m_4\\
-x(\frac{1}{|x^2+y^2|^{3/2}}-\frac{\lambda}{M})m_5+(2+x)(\frac{1}{|x^2+y^2+4x+4|^{3/2}}-\frac{\lambda}{M})m_6=0,
\end{split}
\end{equation}
\begin{equation}
\begin{split}
2y(\frac{1}{|2y|^3}-\frac{\lambda}{M})m_1+0m_2+0m_3+2y(\frac{1}{|4+4y^2|^{3/2}}-\frac{\lambda}{M})m_4\\
+y(\frac{1}{|x^2+y^2|^{3/2}}-\frac{\lambda}{M})m_5+y(\frac{1}{|x^2+y^2+4x+4|^{3/2}}-\frac{\lambda}{M})m_6=0,
\end{split}
\end{equation}
\begin{equation}
\begin{split}
-2(\frac{1}{|4+4y^2|^{3/2}}-\frac{\lambda}{M})m_1-2(\frac{1}{2^3}-\frac{\lambda}{M})m_2+0m_3+0m_4\\
-(2+x)(\frac{1}{|x^2+y^2+4x+4|^{3/2}}-\frac{\lambda}{M})m_5+x(\frac{1}{|x^2+y^2|^{3/2}}-\frac{\lambda}{M})m_6=0,
\end{split}
\end{equation}
\begin{equation}
\begin{split}
2y(\frac{1}{|4+4y^2|^{3/2}}-\frac{\lambda}{M})m_1+0m_2+0m_3+2y(\frac{1}{|2y|^3}-\frac{\lambda}{M})m_4\\
+y(\frac{1}{|x^2+y^2+4x+4|^{3/2}}-\frac{\lambda}{M})m_5+y(\frac{1}{|x^2+y^2|^{3/2}}-\frac{\lambda}{M})m_6=0,
\end{split}
\end{equation}
\begin{equation}
\begin{split}
-2(\frac{1}{2^3}-\frac{\lambda}{M})m_1-2(\frac{1}{|4+4y^2|^{3/2}}-\frac{\lambda}{M})m_2+0m_3+0m_4\\
-(2+x)(\frac{1}{|x^2+y^2+4x+4|^{3/2}}-\frac{\lambda}{M})m_5+x(\frac{1}{|x^2+y^2|^{3/2}}-\frac{\lambda}{M})m_6=0,
\end{split}
\end{equation}
\begin{equation}
\begin{split}
0m_1-2y(\frac{1}{|4+4y^2|^{3/2}}-\frac{\lambda}{M})m_2-2y(\frac{1}{|2y|^3}-\frac{\lambda}{M})m_3+0m_4\\
-y(\frac{1}{|x^2+y^2+4x+4|^{3/2}}-\frac{\lambda}{M})m_5-y(\frac{1}{|x^2+y^2|^{3/2}}-\frac{\lambda}{M})m_6=0,
\end{split}
\end{equation}
\begin{equation}
\begin{split}
x(\frac{1}{|x^2+y^2|^{3/2}}-\frac{\lambda}{M})(m_1+m_2)+\\
(x+2)(\frac{1}{|x^2+y^2+4x+4|^{3/2}}-\frac{\lambda}{M})(m_3+m_4)\\
+0m_5+2(1+x)(\frac{1}{|2(1+x)|^3}-\frac{\lambda}{M})m_6=0,
\end{split}
\end{equation}
\begin{equation}
\begin{split}
(x+2)(\frac{1}{|x^2+y^2+4x+4|^{3/2}}-\frac{\lambda}{M})(m_1+m_2)\\
+x(\frac{1}{|x^2+y^2|^{3/2}}-\frac{\lambda}{M})(m_3+m_4)+\\
+2(1+x)(\frac{1}{|2(1+x)|^3}-\frac{\lambda}{M})m_5+0m_6=0,
\end{split}
\end{equation}
By(2.3),(2.5),(2.7) and (2.9),we have:
\begin{equation}
\begin{split}
(m_3-m_4)(\frac{1}{|4+4y^2|^{3/2}}-\frac{1}{2^3})=0,\\
(m_1-m_2)(\frac{1}{|4+4y^2|^{3/2}}-\frac{1}{2^3})=0.
\end{split}
\end{equation}
By(2.4),(2.6),(2.8) and (2.10),we have:
\begin{equation}
\begin{split}
(\frac{1}{|4+4y^2|^{3/2}}-\frac{1}{2^3})(m_1-m_3)+(\frac{1}{|2y|^3}-\frac{1}{2^3})(m_4-m_2)=0,\\
(\frac{1}{|2y|^3}-\frac{1}{2^3})(m_1-m_3)+(\frac{1}{|4+4y^2|^{3/2}}-\frac{1}{2^3})(m_4-m_2)=0.
\end{split}
\end{equation}
By(2.13) and (2.14),we have:
\begin{equation}
m_1=m_2=m_3=m_4.
\end{equation}
By (2.4),(2.6),(2.11), (2.12) and (2.15),we have
\begin{equation}
m_5=m_6.
\end{equation}
The proof of \textbf{Theorem1.1} is completed.
\subsection{The Proof of Theorem 1.2}
Notice that $(q_1,\cdots,q_6)$ is a central configuration if and only if
\begin{equation}
\sum_{j=1,j\neq k}^{6}\frac{m_{j}m_{k}}{|q_{j}-q_{k}|^{3}}(q_{j}-q_{k})=-\lambda m_{k}q_{k},1\leqslant k\leqslant 6.
\end{equation}
Since the symmetries,(2.17) is equivalent to
\begin{equation}
\sum_{j=1,j\neq k}^{6}\frac{m_{j}m_{k}}{|q_{j}-q_{k}|^{3}}(q_{j}-q_{k})=-\lambda m_{k}q_{k}, k=2,5.
\end{equation}
Now (2.18) is equivalent to
\begin{equation}
\lambda=\frac{1}{4}+\frac{1}{4|1+y^2|^{3/2}}-\frac{xm}{|x^2+y^2|^{3/2}}+
\frac{(2+x)m}{|x^2+y^2+4x+4|^{3/2}},
\end{equation}
\begin{equation}
\lambda=\frac{1}{4y^3}+\frac{1}{4|1+y^2|^{3/2}}+\frac{m}{|x^2+y^2|^{3/2}}+
\frac{m}{|x^2+y^2+4x+4|^{3/2}},
\end{equation}
\begin{equation}
\lambda=\frac{2x}{|x^2+y^2|^{3/2}(1+x)}+
\frac{2(2+x)}{|x^2+y^2+4x+4|^{3/2}(1+x)}+\frac{m}{4|1+x|^3},
\end{equation}
(2.19) ,(2.20) and (2.21) are equivalent to
\begin{equation}
(\frac{1+x}{|x^2+y^2|^{3/2}}-\frac{1+x}{|x^2+y^2+4x+4|^{3/2}})m=\frac{1}{4}(1-\frac{1}{y^3}),
\end{equation}
\begin{equation}
\begin{split}
(\frac{1}{|x^2+y^2|^{3/2}}+\frac{1}{|x^2+y^2+4x+4|^{3/2}}-\frac{1}{4|1+x|^3})m=\\ \frac{2x}{|x^2+y^2|^{3/2}(1+x)}+
\frac{2(2+x)}{|x^2+y^2+4x+4|^{3/2}(1+x)}-\frac{1}{4y^3}-\frac{1}{4|1+y^2|^{3/2}},
\end{split}
\end{equation}
By (2.22) we have
\begin{equation}
m=m_1(x,y)=\frac{1}{4}(1-\frac{1}{y^3})\frac{|x^2+y^2|^{3/2}|x^2+y^2+4x+4|^{3/2}}{(1+x)(|x^2+y^2+4x+4|^{3/2}-|x^2+y^2|^{3/2})}
\end{equation}
$m_1(x,y)>0$ if and only if $y>1$.\\
By (2.23) we have
\begin{equation}
\begin{split}
m=m_2(x,y)=
[\frac{2x}{|x^2+y^2|^{3/2}(1+x)}+
\frac{2(2+x)}{|x^2+y^2+4x+4|^{3/2}(1+x)}\\-\frac{1}{4y^3}-\frac{1}{4|1+y^2|^{3/2}}]
\times [\frac{1}{|x^2+y^2|^{3/2}}+\frac{1}{|x^2+y^2+4x+4|^{3/2}}-\frac{1}{4|1+x|^3}]^{-1}
\end{split}
\end{equation}
Then $(q_1,\cdots,q_6)$ is a central configuration if and only if
\begin{equation}
m_1(x,y)=m_2(x,y)>0
\end{equation}
It is obvious that \\
\begin{equation}
m_1(1,\sqrt{3})=m_2(1,\sqrt{3})=1.
\end{equation}
\begin{equation}
\frac{\partial m_1(1,\sqrt{3})}{\partial x}=\frac{1}{4}.
\end{equation}
\begin{equation}
\frac{\partial m_2(1,\sqrt{3})}{\partial x}=\frac{1}{2}\frac{(9-16\sqrt{3})}{(27+4\sqrt{3})}\neq\frac{1}{4}.
\end{equation}
By implicit function theorem,there exists exists a non-empty open set $U$ and
$\varphi(y)\in C(U)$ such that,$\sqrt{3}\in U$ , $\varphi(\sqrt{3})=1$ and  $\forall y \in U$,
$m_1(\varphi(y),y)=m_2(\varphi(y),y)$.
\\The proof of \textbf{Theorem1.2} is completed.

\end{document}